# Investigating behavior change indicators and cognitive measures in persuasive health games

Durga, S., Hallinan, S., Seif El-Nasr, M., Shiyko, M., and Sceppa, C

**ABSTRACT**
Outcome-driven studies designed to evaluate potential effects of games and apps designed to promote healthy eating and exercising remain limited either targeting design or usability factors while omitting out health-based outcomes altogether, or tend to be too narrowly focuses on behavioral outcomes within a short periods of time thereby less likely to influence longitudinal factors that can help sustain healthy habits. In this paper we argue for a unified approach to tackle behavioral change through focusing on both health outcomes and *cognitive precursors*, such as players' attitudes and behaviors around healthy eating and exercising, motivation stage and knowledge and awareness about nutrition or physical activity. Key findings from a 3-month long game play study, with 47 female participants indicate that there are clear shifts in players' perceptions about health and knowledge about eating. This paper extends our current understandings about approaches for evaluating health games and presents a unified approach to assess effectiveness of game-based health interventions through combining health-based outcomes and shifts in players' cognitive precursors.

**Author Keywords**
health-based games, games for health, persuasive games, adherence, body mass index, readiness to change

**ACM Classification Keywords**
H5.m. Information interfaces and presentation (e.g., HCI): Miscellaneous.

## General Terms
Persuasive games, health, Transtheoretical model, health behavior change.

## INTRODUCTION
In the past decade or so, there has been a constant rise of use of pervasive and persuasive mobile technologies, applications and games that can promote healthy behaviors. There is a growing interest to design innovative technological interventions that can promote long-term solutions while also being accessible and well integrated with individuals' life style, which can be seen in the sheer prevalence of health and wellness mobile apps and games that seek to "gamify" experiences with physical activity and healthy eating. Pew reports that as of 2014, at least 72 % of cell phone users went online seeking for health information about diseases or certain conditions, treatments or procedures, and that 7 out of 10 in the U.S, have tracked a health indicator for themselves through mobile or online apps. Because of such ubiquity of access to health technologies, research initiatives are increasingly geared towards designing health technologies that can incentivize and sustain health behavior change in the long-term. Research in this area continues to be critical due to its potential implications for longitudinal outcomes in health, such as combating obesity and other chronic ailments caused by unhealthy weight gain, a pressing issue that has persisted over the last two decades [1], [2].

However, outcome-oriented studies, for the most part, have remained limited in establishing the factors that can successfully demonstrate behavioral outcomes in the long-term. Most outcome studies tend be short-term or within one or two play session in efforts to measure usability factors, like influence of avatar designs on body image and perceptions on self-efficacy [3], or pre- and post- gains in intensity of physical activity within a play session [4]. In addition, often these types of work are conducted within lab-settings in order to assure consistency with experimental conditions and control them. Such methods have recently come under serious critique by researchers in the field of HCI who point to the inherently ubiquitous and "accessible-anywhere" properties of health technologies and suggest that efficacy measures for health technologies ought to focus on specific intervention process-oriented strategies [5], such as encouraging certain behaviors through leveraging lifestyle patterns and "encroaching upon" individuals' personal as well as their social worlds [6]. Other similar works argue for incentivizing health behavior change based on personalized design features catering to different personality types [7], [8]. In other words, more recent work increasingly seek to address health behavior change using models of *acceptance* and *impact* of health technologies on individuals' behaviors [9].

To sum up, a critical area of challenge in designing outcomes-based studies to evaluate efficacy of health technologies is that these studies have either a) tended to target design and usability factors thereby omitting out on health-based outcomes, or b) tend to be too narrowly focused on behavioral outcomes. Contemporary outcome-based research in health technologies, we argue, ought to consider not only health-based outcomes (adherence to physical activity, gain or loss of weight etc), but also monitor for shifts in cognitive precursors to specific behavioral measures, such as self-efficacy, change in attitudes, or perception, motivation stage and knowledge and awareness about nutrition or physical activity [10].

Thus, with this paper, we specifically seek to address these aspects through investigating the effects of longitudinal gameplay on behaviors *in real life* and to observe shifts, if any, in terms of player attitudes or cognitive shifts and behavioral outcomes. In this paper, we present results from a 3-month long game-play study that we ran with 47 participants in which players were asked to play the game, as well as respond to monthly questionnaires assessing participants' weight loss or gain, *stage of health behavior change* and *nutritional knowledge*. Findings from the study indicate that there are clear shifts in players' perceptions about health and knowledge about eating. The current study contributes towards extending our existing understanding on how to go about measuring longitudinal outcomes in health behaviors and persuasive technologies and presents recommendations towards modification to the design of the study to pursue these goals, particularly in designing studies that seek to measure cognitive indicators complimenting behavioral outcomes.

This paper intends to extend findings from our earlier investigations conducted on the same health game to tackle design issues pertinent to player engagement and health behavior change (removed for blind review). The paper is divided as follows. We first discuss previous work in the area of health behavior change and games or ubiquitous technology. We then discuss the game outlining the design. We then discuss a longitudinal study we conducted to investigate the impact of the game on health behavior change. We present results and discuss the impact of these results. The results and discussion constitute the contribution of this paper addressing the challenge, alluded to above, regarding the separation of behavioral outcomes and cognitive measures in measuring effectiveness of health interventions.

## PREVIOUS WORK

### Studies measuring impact of exergames on health behavior change through outcomes

For the most part, the range of games designed to promote healthy behaviors include exergames, or games that incorporate motion-sensing devices that help to track movement and thereby motivate players to exert themselves physically. In these games, users usually see a simulated virtual representation of themselves (Avatar) or part of themselves (tracked limbs) while they are asked to perform a task by moving their body. Such games are designed to be physically intuitive and suitable for users with not much prior gaming experience. Popular examples include Nintendo Wii Fit, which facilitates diet and exercise tracking on a game console, Wii Sports, which includes physical movement as part of gameplay, Zumba game on Kinect, which incorporates dance to motivate exercise and My Weight Loss Coach and the DS game Pokémon HeartGold, which use activity-tracking monitors, like pedometers, to promote real-life physical activities. Overall, in exergames the design premise is to use avatar representation and visual display of ongoing real-time activity on screen to engage players in physical activity for a long play session. Such studies investigating health effects on playing exergames are limited to short-term or more immediate player response to the game and measure physical exertion in a one-time or a pre-and-post gameplay sessions [3], [4], [11]. In the next section, we present a few examples from studies that have sought to tackle behavior change indicators using cognitive measures in addition to immediate physiological outcomes.

### Studies measuring impact of games and apps on health behavior change using an integrated behavior model and cognitive measures

More recently, studies of persuasive games and ubiquitous health technologies have started to investigate health-behavior change by designing game environments that can transcend "console-play". These studies have looked at design techniques that can strategically empower individuals with relevant information regarding health using compelling game narratives [12], generating awareness about certain ailments and how to manage them [13] or educating players about healthy nutritional choices [14]. Such studies leverage rewards to incentivize some of the activities that individuals tended to engage in more frequently when presented a different technology either to monitor their activity through devices, like mobile activity trackers [7], or through providing small goals with progressive rewards mechanisms [6][10]. Similar conclusions have also been presented by studies of ubiquitous health technologies, wherein researchers found that *persuasive* design factors in a health technology explicitly promotes certain behaviors through its design, while the study remains unable to draw causal relationship between device use and health and behavior effects [15]. This is an appropriate example illustrating the challenge of designing outcome-based studies in health.

Within the health sciences, for several years now, research on longitudinal behavior change have successfully demonstrated and empirically verified several integrated behavior models that take both *action* and *process* into account while designing successful health interventions. These include frameworks such as transtheoretical model of behavior change [16], precaution adoption process model [17] and planned reasoned action, planned behavior models [18]. Prochaska and Velicer's extensive review of health interventions based on the transtheoretical model is a great example on how integrated outlook on behavior change influences effectiveness of interventions, and therefore necessitates both behavioral and cognitive measures [16]. In it, the authors state that when interventions are personalized, or "stage-matched", they are able to produce remarkable impacts on a majority of at-risk populations. In the following sections we describe the context of the study, rationale for using specific evaluative cognitive measures targeting movement in cognitive precursors (e.g. motivation stage) while corroborating it to health-based outcomes (e.g. weight loss).

**STUDY CONTEXT: A HEALTH AND WEIGHT-LOSS GAME**

This paper is based on an empirical investigation of long-term impact of playing an online social game designed to incentivize healthy eating and exercising habits, eventually promoting weight-loss in adults, particularly women. The data corpus for the study and findings presented in this paper are drawn from an ongoing, longitudinal initiative to study the potential of persuasive gaming technologies to promote positive health behavior change.

Through an ongoing design and research collaboration with a commercial game company [removed for blind review], we embedded several design strategies within the health game to incentivize real-life behavior change. The central design premise of the game is to bring playfulness to physical activities that can be incorporated into a daily routine and reward players accordingly for the efforts they make to be physically active in their lives *outside* of the game. The game, titled *SpaPlay* (see figure 1), is a health spa island, in which each user starts with a few buildings or resources, an avatar, and some virtual currency. *SpaPlay*, like many other social media environments, operates under an economy model, which encourages the collection of virtual currency to allow players to purchase tangible and intangible rewards. Tangible rewards may be coupons, or invitations to exclusive deals from partner companies. Intangible or virtual rewards are power-ups in the games within *SpaPlay*, or achievement points in the social media environment, which allow players to unlock specific decors to decorate and personalize their environments, or clothing for their avatars. Players can also invite friends and build their social network. The game provides several ways for players to incrementally increase exercising and start eating healthier, through managing their virtual island by doing game *sparks*, which are small activities that players can fit in their everyday routine, like "eat a fruit for snack". The game also lets players plan longer activities, or *quests*, which are more goal-oriented that take about a week to be completed, such as "beginner training for biking". Players also manage their virtual *island-activities*, e.g., picking up trash or adding new vegetation.

The game was designed based on the PENS (Player Experience of Need Satisfaction) model [19] [20] through incorporating specific gamification techniques that engage participants in a virtual game world with rewards and incentives for healthy behaviors outside the game world. SpaPlay interweaves health concepts within the structure of a social virtual world. Drawing upon the three levers of intrinsic motivation as postulated by the theory of self-determination [21] *autonomy, competence* and *relatedness*, the game mechanics in SpaPlay are structured in ways that optimally satisfy player autonomy and give players ways to grow their abilities and gain mastery of through progressively challenging quests that are personally meaningful for meeting their health goals.

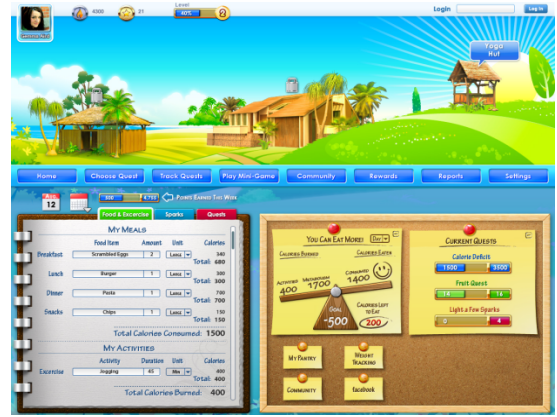

**Figure 1. Tracking of Meals and Calorie Intake. The screenshot also shows, in the upper left, the avatar of the character with virtual currency, levels accomplished, etc.**

For instance, to satisfy player autonomy and competence, *SpaPlay* allows players to track activities (see Figure 1), such as eating, exercising, etc., through several sensory and self report systems. Users can log their eating and food choices through the game. Users can also use physical activity sensors, such as Fitbit, Reflex, etc. These sensors as well as some mobile applications are integrated with the game. All data from these systems are integrated into the game and users can view these data through the game and can get rewards based on these data. *SpaPlay* also uses these data for motivational purposes:

- *Visual feedback and rewards for physical healthy actions*: Character level and appearance are tied to data collection and physical activity; thus rewarding these behaviors with game content changes.
- *Social Dimension to Sharing of Healthy Activities.* Due to its social nature, *SpaPlay* allows users the choice of sharing some information with their friends. This adds an important motivational factor on the relatedness scale from Rigby and Ryan (2011). This is currently implemented in *SpaPlay* in various ways to tie into several motivational concepts discussed in the PENS model. One example is through visualization of activity directed to a targeted challenge, thus appealing to players motivated by competition.

*SpaPlay* also includes several systems for raising awareness to health-based concepts. For example, it includes visualizations of different food types compared to the level of exercise needed. Another example is the use of games that have health-based themes, such as the Yoga game, which embeds understanding of Yoga poses through a bejeweled style of game play, and the Chef game, wherein users play a diner-dash game to put together a healthy dish based on healthy recipes, either encoded in the system or supplied by friends of the user through the social media. It is important to note that while the specifics of the implementation, theme and visualizations differentiate *SpaPlay* from other health based behavior change games, *SpaPlay* shares many of the similar techniques and reward systems to other health-based apps targeting increased physical activity and healthier nutrition choices, like MyFitnessPal, or NutritionHub. Thus, we are not presenting SpaPlay as a unique game, but rather a game that includes many gamification techniques used by other ubiquitous health applications which makes it suitable for study investigating the relationship between its design and various important health measures, such as engagement, adherence, and behavior change.

In other words, while several such games, seeking to incentivize longitudinal health behavior change through similar gamified approaches, often mere design iterations remain limited in bringing about sustained increase in physical activity or adhering to healthy eating habits. The reason for lack of success, we argue, is often beyond specific design implementations and depends upon both behavioral indicators and cognitive factors [22], [23]. Presenting an integrated model for behavior change, Prochaska and Clemente have argued that behavioral processes alone may not predict long-term change and at times, effective interventions are the ones that allow for "*cognitive re-structuring* in the individual's sense of self-efficacy" (p-277, [23]). In this paper, thus, we present shifts in players' cognitive perceptions about health and about themselves and behavioral indicators tracked through in their gameplay, such as changes in weight, amount of exercise done and consistency, as they play SpaPlay for 3 months.

**STUDY DESIGN AND DATA COLLECTION METHODS**
The study design and methodological framework for this paper chiefly aims to understand and investigate the effect of playing the game on both behavioral indicators and cognitive factors. As argued earlier, health behavior change depends not only on observed behavioral outcomes, but also depends upon several precursors, such as self-efficacy [24], nutrition knowledge [25] and barrier to exercise [26], [27]. Accordingly, in order for us to investigate health behavior change, we wanted to both factor in the behavioral output from logs of play actions in the game, while also drawing from player responses on precursors like readiness to change [24], [28] and baseline measures on participants' current physical activity levels, eating preferences and basic overview of their gaming preferences (e.g. console games versus, Cellphone or Facebook Social games) while they participated in the study for 3 months.

**Screening and Participant Recruitment**
For this study we used several screening measures to select participants for the study. In addition to selecting only female participants, since the game is designed for a predominantly female audience [removed for blind review], we also filtered participants based on their motivational *stage of change*.

*Stages of change* is an useful framework for studying behavior change which has been extensively used for designing and tailoring interventions to help individuals manage chronic health management or combat substance abuse. The model posits that in terms of motivation, individuals may be grouped under different stages of *readiness for change* (pre-contemplation, contemplation, preparation, action or maintenance) and given a baseline stage, the model can be used to predict behavior change [24], [29]. The foundational work on readiness for change, or also known as the *transtheoretical model* (TTM), defines these stages as follows: 1) pre-contemplation: no intention to take action within the next 6 months, 2) contemplation: intends to take action within the next 6 months, 3) preparation: intends to take action within the next 30 days and has taken some behavioral steps in this direction, 4) action: changed overt behavior for less than 6 months, and 5) maintenance: changed overt behavior for more than 6 months [30]. The rationale behind detecting the motivational grouping of individuals is that one might design interventions that would promote certain decisions that are more likely to empower individuals to move through to the next stage in making positive change in behavior.

As posited earlier, since we deem individuals' predispositions towards health and their intrinsic motivation to be precursors to sustained adherence and health behavior change, we screened participants using the *Readiness to change questionnaire* recruiting individuals in the contemplation, or preparation stages to detect for shifts in their perceptions due to long-term gameplay. In addition, gains in knowledge about nutrition and healthy eating and body weight differences over the period of 3 months were also collected; once every three weeks. We asked participants to fill in the questionnaires and BMI measures at baseline and once every month, resulting in four data points during the 3-month study.

Thus, the data corpus of the study includes:

1) behavioral logs automatically collected and time stamped for every player action in the game . These records captured all participants' engagement with the game within a day across the 3 months of the study. It should be noted that behavioral logs here are only player actions in game and do not show or uncover motivational or cognitive elements for why people exhibited such behaviors.

2) responses to a 94-item questionnaire, composed of: the TTM/stages of change constructs, nutritional knowledge

and baseline body weight measures [25], [28]. The instrument was administered at baseline (beginning of the study) and every 4 weeks in the study for the period of 3 months.

In addition, information pertinent to participant demographics were also collected during the first screening that included age, highest education level attained, ethnicity, weight and height. In addition, we collected data on eating and physical activity behaviors as well as motivation to change these behaviors.

## RESULTS

### Summative Overview of Participant Breakdown based on Baseline Measures

Of the 60 female participants initially recruited, who were screened to be in the pre-contemplation or contemplation change based on their responses to the readiness to change questionnaire (explained in further detail in previous section), 13 dropped out within the first week of the study. Primary reasons that participants attributed to dropping out included:

1. Lack of gaming experience or unfamiliarity with gaming — these were older women (over 50) — and reported lack of interest in games.

2. Lack of commitment and inability to provide adequate time to the study — again, these were also individuals who were in pre-contemplation stage. It is to be noted that while the stages are clear constructs to detect individuals' motivational levels, as Prochaska, Redding & Evers caution, definitions of stages are anything but mutually exclusive, especially when "health behaviors are complex" (p-142)[30]. This means individuals in lower stages would need to perceive a greater number of benefits to adopting a certain behavior or else easily fall prey to minimal barriers to participation [16].

As Table 1 demonstrates, of the 47 participants who were in the study the majority were overweight or obese, predominantly reporting low regularity in high intensity physical activity. Participants' mean baseline BMI was 27 (md = 25.54, sd = 5.49). Mean age was 29.76 years old (md = 28, sd = 7.33). In terms of gaming behaviors, the majority reported cell phones to be their preferred gaming device, and most of them reported playing at least 4-5 times a week. They also preferred social Facebook or mobile games. The preliminary snapshot of data indicates that the sample was largely a likely audience for a game like SpaPlay, which incorporates several of the gaming preferences reported by the participants, such as casual gaming on a mobile/social media platform.

**Table 1. Overview of participant breakdown based on their baselines (n=47)**

| | |
|---|---|
| 1. Education level | 52% Undergraduate |
| | 29% Masters |
| | 19 % High-school or equivalent |
| 2. Ethnicity | 65% White |
| | 18% Asian |
| | 8% Black |
| | 4% Hispanic |
| | 5% Multiracial |
| 3. Baseline BMI | 15% Normal |
| | 30% Overweight |
| | 33% Obese |
| | 9% Underweight |
| 4. Frequency of intense physical activity | 41% Hardly ever |
| | 26% At most once a week |
| | 22% 3-4 times weekly |
| | 11% Regularly/daily |
| 5. Preferred gaming platform | 75% Cellphone or iPod touch only |
| | 21% Consoles (Xbox, PlayStation etc.) |
| | 4% PC |
| 6. Favorite genre of games | 43% Facebook games |
| | 40% Mobile games |
| | 14 % MMORPG |
| | 3% FPS |
| 7. Frequency of gaming reported | 52% At least 4-5 times a week |
| | 22 % 2-3 times a week |
| | 22% At least once a week |
| | 4% Never |

### Descriptive Statistics for Adherence to Game

Using real-time telemetry data from the game, adherence was quantified in three ways. First, we computed the number of days (out of total 90 possible) participants engaged with the game. Any real-time records of log-ins on a given day were used as indicators of engagement. Data showed that participant's mean number of logins was 8.57 (md=2, sd=12.5).

Second, to assess regularity of adherence to game play, we computed time intervals between successive days of play for each participant. Mean days between logins served as a measure of the length of time that elapsed between participants' use of the game, an indicator of negative adherence. A summary of all participants' mean length between logins shows mean 24.7 days between logins (md=22, sd=14.23). Standard deviation of days between logins was computed for each individual, and served as a measure of regularity of play. Higher standard deviation indicated highly irregular usage, lower standard deviation indicates more regular usage. Average standard deviation of length between logins for participants was 16.48 (md = 15.16, sd=7.66). Overall, this shows varied but low level of adherence to the game among our participants.

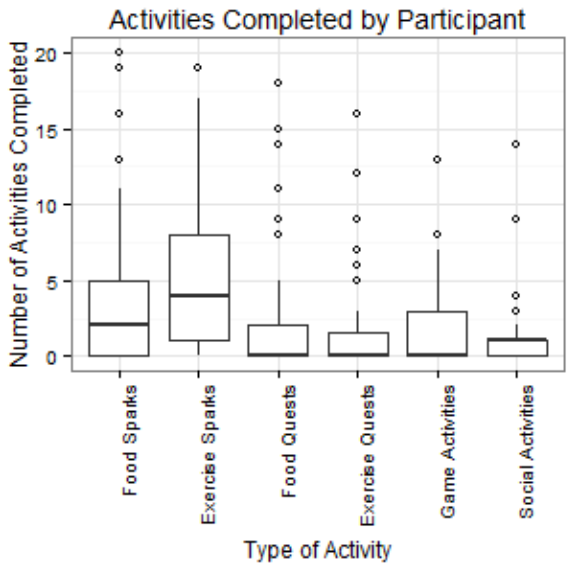

**Figure 2. Shows summaries of the data collected on activities completed by participants within the game, across the 6 categories available.**

Figure 2 shows summary data of gameplay activities with variance. As the figure shows, participants showed a preference for Spark activities. But there is a high level of variability in program utilization as shown in the figure.

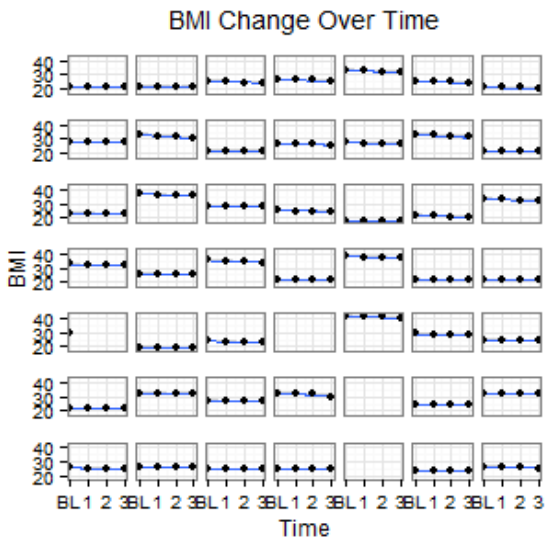

**Figure 3. HLM Results for BMI**

**Changes in Health and Motivation Outcomes**
Changes in the outcome variables were observed during the course of the study. In particular, there is a significant change in BMI, TTM stage, and Nutrition Knowledge.

Change in participants' BMI was shown in almost all cases, as demonstrated in Figure 3. The figure shows the measure of BMI at BaseLine (BL), 1st, 2nd and 3rd measurement time (as discussed above, we asked participants to fill in the questionnaires and BMI measures at baseline and once every month). A one-tailed paired sample's t-test confirms a significant change in BMI, $t(44) = -7.7$, $p <0.01$. Cohen's d effect size of 1.07 indicates a large decrease in BMI occurred in the course of the study.

For Nutrition Knowledge, one-tailed paired samples t-test produces $t(44)=4.28$, $p<0.01$, and Cohen's d of 0.83 – a large increase in nutrition knowledge.

As TTM stage changes were non-parametric, Wilcoxon signed rank tests for matched pairs were used. Significant changes were shown in all domains. For Healthy Eating, $V= 338$, $p<0.01$; for Physical Activity, $V= 351$, $p<0.01$; and for Beverage Choices $V=630$, $p<0.01$.

Furthermore, we used generalized multilevel linear modeling to determine whether or not the changes indicated above were linked to gameplay. For this we used the game play data and looked at correlations between outcomes and gameplay data over time. Figures 4 and 5 show these results for Nutrition Knowledge and BMI.

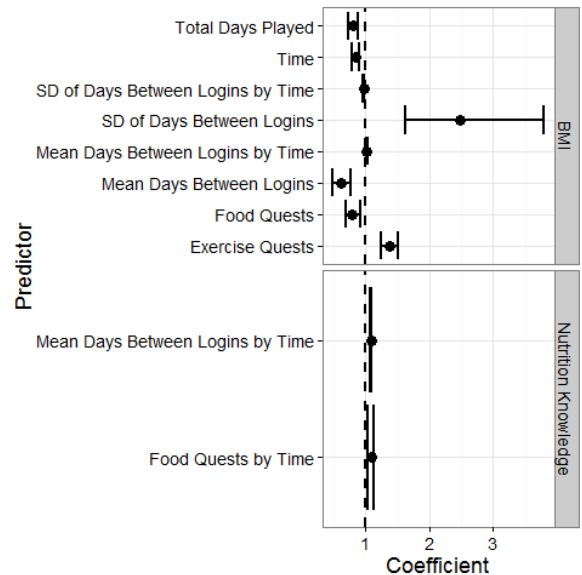

**Figure 4. HLM Results for Nutrition Knowledge**

The included predictors were all variables that were statistically significant for exploratory analysis ($p < 0.1$) and comprised the final model. The line down the chart (figures 4 and 5) represents the distinction between a positive correlation (right of the line) and a negative correlation (left). The distance of the point from the line represents the coefficient (i.e. the strength of the relationship of the predictor to the outcome), and the error bars represent standard error.

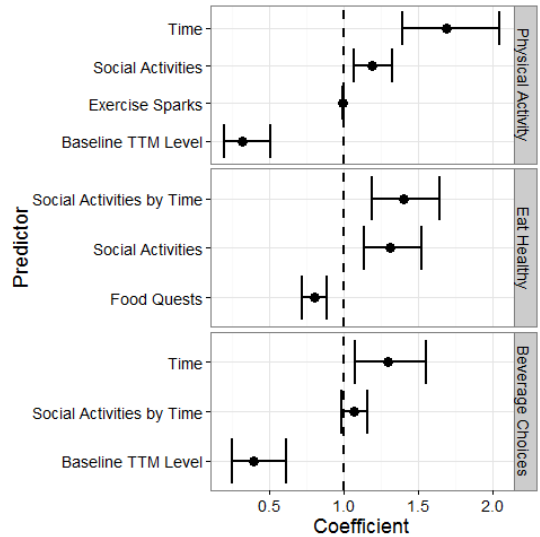

**Figure 5. HLM Results for TTM Stage Change**

In examining the figures, several gameplay activities emerge as significant predictors of all examined outcomes. BMI change was most significantly predicted by SD of days between logins. As negative BMI change indicates weight loss, greater irregularity of game usage was predictive of less weight loss. The results also demonstrate a significant positive effect of Social Activities predicting TTM stage change in physical activity and healthy eating, and a significant interaction effect for social activities and time in beverage choices. These results indicate that social activities within the game predicted a greater likelihood for forward momentum through the TTM stages in the listed domains.

## DISCUSSION

In continuation from above, overall we noticed improvements in indicators primarily external to the game. Results from the analysis indicate that there is shift in TTM stages, NKQ measures and BMI. In other words, as anticipated, and discussed extensively in the literature review sections of this paper, we noticed shifts in cognitive precursors, like their motivation state change, awareness to eating healthy and overall improvements in BMI within the 3 months of the study. Additionally, specific factors from the game were demonstrated to influence these health measures. However, overall adherence and utilization of the program was low, and methods for encouraging users to engage with the game remains a question to be explored.

As has been argued, behavior change is complex and extremely hard to achieve, particularly in the context of longitudinal health management [10] (see also [30] on stages of change and how individuals move along them). However, by taking apart specific behavioral outcomes, such as physical activity or food frequency measures from precursors that are likely to promote interventions, such as cognitive perception on health goals, long-term patterns in activity, knowledge and awareness and self-efficacy measures, we are better positioned to strategically predict certain outcomes (for example, dropping out possibilities), or focus on customizing follow-ups through facilitating individuals to cognitively and behaviorally to advance from one motivational stage to another [30].

## CONCLUSION AND IMPICATIONS FOR FUTURE WORK

For the future, we seek to pursue explanations based on demographic factors and behaviors assessed at baseline to elicit lower adherence values in specific gameplay aspects. Nevertheless, it seems that through playing the game participants' cognitive perceptions or attitudes in terms of drawing attention to what they were eating and how they were exercising changed, in addition to shifts in BMI measures and nutritional knowledge. We wish to highlight that positive changes in health-behaviors and attitudes are crucial first steps towards longitudinal success of any health intervention, especially since it can be challenging to observe these changes in the short term. As a next step in continuing this work, we seek to examine relationships between specific game activity types to cognitive and behavioral outcomes, based on specific sub-groups, such as baseline measures (overweight, obese or normal weight), and demographic measures (education level and ethnicity).

## ACKNOWLEDGMENTS

Removed for blind review.

## REFERENCES

[1] E. A. Finkelstein, J. G. Trogdon, J. W. Cohen, and W. Dietz, "Annual medical spending attributable to obesity: payer-and service-specific estimates," *Health Aff.*, vol. 28, no. 5, pp. w822–w831, 2009.

[2] and E. J. S. Sebelius, Kathleen, Thomas R. Frieden, "Health, United States, 2011: With special feature on socioeconomic status and health," 2012.

[3] H. Song, W. Peng, and K. M. Lee, "Promoting Exercise Self-Efficacy with an Exergame," *J. Health Commun.*, vol. 148, 2011.

[4] K. Jacobs, L. Zhu, M. Dawes, J. Franco, A. Huggins, C. Igari, B. Ranta, and A. Umez-Eronini, "Wii health: a preliminary study of the health and wellness benefits of Wii Fit on university students," *Br. J. Occup. Ther.*, vol. 74, no. 6, p. 7, 2011.

[5] P. Klasnja, S. Consolvo, and W. Pratt, "How to evaluate technologies for health behavior change in HCI research," in *Proceedings of the SIGCHI Conference on Human Factors in Computing Systems*, 2011, pp. 3063–3072.

[6] S. Consolvo, D. W. McDonald, and J. A. Landay, "Theory-driven design strategies for technologies that support behavior change in everyday life," in

*Proceedings of the SIGCHI Conference on Human Factors in Computing Systems*, 2009, pp. 405–414.
[7] S. M. Arteaga, M. Kudeki, and A. Woodworth, "Combating obesity trends in teenagers through persuasive mobile technology," *ACM SIGACCESS Access. Comput.*, no. 94, pp. 17–25, 2009.
[8] S. M. Arteaga, M. Kudeki, A. Woodworth, and S. Kurniawan, "Mobile system to motivate teenagers' physical activity," in *Proceedings of the 9th International Conference on Interaction Design and Children*, 2010, pp. 1–10.
[9] S. Halko and J. A. Kientz, "Personality and persuasive technology: an exploratory study on health-promoting mobile applications," in *Persuasive technology*, Springer, 2010, pp. 150–161.
[10] P. Klasnja, S. Consolvo, D. W. McDonald, J. A. Landay, and W. Pratt, "Using mobile & personal sensing technologies to support health behavior change in everyday life: lessons learned," in *AMIA Annual Symposium Proceedings*, 2009, vol. 2009, p. 338.
[11] C. O'Donovan and J. Hussey, "Active video games as a form of exercise and the effect of gaming experience: a preliminary study in healthy young adults," *Physiotherapy*, vol. 98, no. 3, pp. 205–210, 2012.
[12] L. Gillis, "Use of an interactive game to increase food acceptance--a pilot study.," *Child. Care. Health Dev.*, vol. 29, no. 5, pp. 373–5, Sep. 2003.
[13] N. Aoki, S. Ohta, H. Masuda, T. Naito, T. Sawai, K. Nishida, T. Okada, M. Oishi, Y. Iwasawa, K. Toyomasu, K. Hira, and T. Fukui, "Edutainment tools for initial education of type-1 diabetes mellitus: initial diabetes education with fun.," *Stud. Health Technol. Inform.*, vol. 107, no. Pt 2, pp. 855–9, Jan. 2004.
[14] T. Baranowski, J. Baranowski, K. W. Cullen, T. Marsh, N. Islam, I. Zakeri, L. Honess-Morreale, and C. demoor, "Squire's Quest!: Dietary outcome evaluation of a multimedia game," *Am. J. Prev. Med.*, vol. 24, no. 1, pp. 52–61, 2003.
[15] T. Fritz, E. M. Huang, G. C. Murphy, and T. Zimmermann, "Persuasive technology in the real world: a study of long-term use of activity sensing devices for fitness," in *Proceedings of the 32nd annual ACM conference on Human factors in computing systems*, 2014, pp. 487–496.
[16] J. O. Prochaska and W. F. Velicer, "The transtheoretical model of health behavior change," *Am. J. Heal. Promot.*, vol. 12, no. 1, pp. 38–48, 1997.
[17] N. D. Weinstein and P. M. Sandman, "A model of the precaution adoption process: evidence from home radon testing.," *Heal. Psychol.*, vol. 11, no. 3, p. 170, 1992.
[18] D. E. Montano and D. Kasprzyk, "Theory of reasoned action, theory of planned behavior, and the integrated behavioral model," *Heal. Behav. Heal. Educ. Theory, Res. Pract.*, vol. 4, pp. 67–95, 2008.
[19] S. Rigby and R. M. Ryan, *Glued to Games: How Video Games Draw Us In and Hold Us Spellbound: How Video Games Draw Us In and Hold Us Spellbound*. ABC-CLIO, 2011.
[20] S. Rigby and R. Ryan, "The Player Experience of Need Satisfaction (PENS)," 2007.
[21] E. L. Deci and R. M. Ryan, *Intrinsic Motivation and Self-Determination in Human Behavior (Perspectives in Social Psychology)*. Plenum Press, 1985, p. 371.
[22] R. Schwarzer, "Modeling health behavior change: How to predict and modify the adoption and maintenance of health behaviors," *Appl. Psychol.*, vol. 57, no. 1, pp. 1–29, 2008.
[23] J. O. Prochaska and C. C. DiClemente, "Transtheoretical therapy: Toward a more integrative model of change.," *Psychother. Theory, Res. Pract.*, vol. 19, no. 3, p. 276, 1982.
[24] B. H. Marcus, V. C. Selby, R. S. Niaura, and J. S. Rossi, "Self-efficacy and the stages of exercise behavior change," *Res. Q. Exerc. Sport*, vol. 63, no. 1, pp. 60–66, 1992.
[25] K. Parmenter and J. Wardle, "Development of a general nutrition knowledge questionnaire for adults," *Eur. J. Clin. Nutr.*, vol. 53, no. 4, pp. 298–308, 1999.
[26] S. A. Brown, "Measuring perceived benefits and perceived barriers for physical activity," *Am. J. Health Behav.*, vol. 29, no. 2, pp. 107–116, 2005.
[27] B. H. Marcus, W. Rakowski, and J. S. Rossi, "Assessing motivational readiness and decision making for exercise.," *Heal. Psychol.*, vol. 11, no. 4, p. 257, 1992.
[28] A. Steptoe, S. Kerry, E. Rink, and S. Hilton, "The impact of behavioral counseling on stage of change in fat intake, physical activity, and cigarette smoking in adults at increased risk of coronary heart disease.," *Am. J. Public Health*, vol. 91, no. 2, p. 265, 2001.
[29] S. J. Curry, A. R. Kristal, and D. J. Bowen, "An application of the stage model of behavior change to dietary fat reduction," *Health Educ. Res.*, vol. 7, no. 1, pp. 97–105, 1992.
[30] J. O. Prochaska, C. A. Redding, and K. E. Evers, "Transtheoretical Model of Behavior Change," in *Encyclopedia of Behavioral Medicine*, Springer, 2013, pp. 1997–2000.